\bigskip
\documentclass[onecolumn,showpacs,preprintnumbers,amsmath,amssymb]{revtex4}
\usepackage{graphicx}% Include figure files
\usepackage{fancyhdr}
\usepackage{dcolumn}% Align table columns on decimal point
\usepackage{bm}
\usepackage{textcomp}
\begin{document}
\preprint{ }
\title{Relativistic symmetry breaking in light kaonic nuclei}
\author{Rong-Yao Yang$^1$, Wei-Zhou Jiang$^1$, Qian-Fei Xiang$^2$, Dong-Rui Zhang$^1$, Si-Na Wei$^1$}
\affiliation{$^1$Department of Physics, Southeast University,
Nanjing 210000, China\\
$^2$Institute of High Energy Physics, Chinese Academy of Sciences,
Beijing 100049 , China}
%\date{}

{\begin{abstract} As the experimental data from kaonic atoms and
$K^{-}N$ scatterings imply that the $K^{-}$-nucleon interaction is strongly
attractive at saturation density,  there is a possibility to form
$K^{-}$-nuclear bound states or kaonic nuclei. In this work, we
investigate the ground-state properties of the light kaonic nuclei
with the relativistic mean-field theory. It is found that the
strong attraction between $K^{-}$ and nucleons reshapes the scalar
and vector meson fields, leading to the remarkable enhancement of the
nuclear density  in the interior of light kaonic nuclei and the
manifest shift of the single-nucleon energy spectra and magic
numbers therein. As a consequence, the pseudospin symmetry is shown
to be violated together with enlarged spin-orbit splittings in these
kaonic nuclei.
\end{abstract}}

\keywords{ Pseudospin symmetry, spin-orbit splitting, kaonic nuclei,
relativistic mean field theory }
\pacs{13.75.Jz, 21.10.Dr, 21.10.Pc, 21.60.Gx}
\maketitle \baselineskip 20.6pt

\section{Introduction}
The pseudospin symmetry (PSS), manifested by the quasidegeneracy
between single-nucleon states with quantum numbers $(n, l , j = l +
1/2)$ and $(n-1, l+2, j = l+ 3/2)$, was found more than 40 years
ago~\cite{69AA,69KTH}. Subsequently, substantial efforts had been
devoted to understanding the dynamic origin of the PSS until it was
recognized that the PSS is a symmetry hidden in the equations for the
small component of Dirac spinors. This might be natural and became
clear later on since the relativistic mean-field (RMF) models are
characteristic of the dynamical description of the spin degree of
freedom and spin-orbit
interactions~\cite{87JD,92CB,95ALB,97JNG,98JM,99JM,99JNGI}. The exact
PSS is a consequence of the fact that the scalar and vector
potentials are equal in size but opposite in sign, i.e.,
$\Sigma(r)=S(r)+V(r)=0$, while in practice the PSS is approximate
since the condition $\Sigma(r)=0$ gives no bound state as in the RMF
theory.  However, the exception was found in Ref.~\cite{13PA} when
the confining potential exists. In the Schr\"odinger-like equation
for the small component of the Dirac spinor, the term proportional to
${d\Sigma{(r)}}/{dr}$ is related to the breaking of the PSS, and the
small magnitude of that term gives the approximate PSS. For further
understanding the properties of the PSS, people have studied the PSS
in a number of physical processes and phenomena including
superdeformation~\cite{87JD}, nucleon-nucleon and nucleon-nucleus
scatterings~\cite{99JNG,02JNG}, single particle resonant
states~\cite{12BNL}, superheavy magic structures~\cite{13JJL},
 identical bands~\cite{90WN}, and
pseudospin partner bands~\cite{08QX,09WH}, and so on.

The PSS breaking, proportional to the radial gradient of the
potential $\Sigma(r)$, can vary with the isospin and charge of nuclei
through the vector potential entry~\cite{98JM,99JM,01PA,01SM,10RL}.
In the past, the PSS breaking had been studied extensively, and some
very useful tools were also developed to analyze the
breaking~\cite{12BNL,13HZL,14JYG}. In addition to the isospin effects
on the PSS breaking, the variation of the isoscalar potentials can
also become one PSS breaking source, although this variation in
different normal nuclei in the nuclear chart is not significant. In
this work, our attention to the PSS breaking is focused on the exotic
systems, the kaonic nuclei that may feature a characteristic
enhancement of the isoscalar density in the core of nuclei.

By analyzing experimental data of kaonic atoms and $K^{-}N$
scatterings, people have found that $K^{-}N$ interaction is strongly
attractive at saturation density but with an optical potential depth
roughly ranging from 40 to 200
MeV~\cite{93EF,94EF,97CJB,99EF,00AR,00SH,00AB,02YA,05JM,06JM,07EF,12DG}.
The great interest in studying kaonic nuclei has been attracted by
the fact that the same sign of the vector and scalar potentials of
the $K^-$ creates a strong attraction that may lead to high-density
$K^{-}$-nuclear bound states.  In the past, lots of theoretical works
have flourished. For instance, few-body
calculations~\cite{02TY,02YA,07NVS,09AD,13AG}, the RMF and
non-relativistic Skyrme-Hartree-Fock researches
~\cite{05JM,06JM,06XHZ,07DG,13XRZ}, were performed to obtain the
$K^-$ binding energy, the width and ground-state properties of
$K^{-}$-nuclear bound states.  Along with various predictions,
experiments for the $K^-$ bound states have also been progressive
with the construction of new
facilities\cite{04TS,07MA,05MA,10TY,06EO,06VKM,08VKM,13SA,13YI}.

Though the PSS is one of fundamental nuclear properties, it has not
received due attention in kaonic nuclei. Indeed, the PSS has seldom
been examined in kaonic nuclei. Thus, it is the aim of this work to
investigate how the PSS is affected by the strong $K^-$-nucleon
attraction in the RMF theory.  For completeness and comparison, we
will also examine the corresponding effect on spin-orbit splittings
that are associated with the spin symmetry. The paper is organized as
follows. The RMF formalism for kaonic ($K^-$) nuclei is given in
Section~\ref{RMF}, and the pseudospin and spin symmetries in RMF are
briefly manifested in section~\ref{PSS}. The results and discussions
are given in section~\ref{result}, followed by a brief summary in
section~\ref{summary}.

\section{RMF formalism for kaonic nuclei}
\label{RMF}
The relativistic Lagrangian containing
$K^{-}$-nucleon interaction can be written as
\begin{eqnarray} \label{Lag}
\mathcal{L}&=& \bar{\psi}_B[ i\gamma_{\mu} \partial ^{\mu} -
M_{B} + g_{\sigma B}\sigma - g_{\omega B}\gamma_{\mu}\omega^{\mu} -
g_{\rho B}\gamma_{\mu}\tau_3 b_0^{\mu} - e\frac{1+\tau_3}
{2}\gamma_{\mu}A^{\mu}]\psi_B \nonumber\\
& &- \frac{1}{4} F_{\mu \nu} F^{\mu \nu}+\frac{1}{2}m^2_{\omega}
\omega_{\mu} \omega^{\mu}
 -  \frac{1}{4} B_{\mu \nu} B^{\mu \nu}+\frac{1}{2}m^2_{\rho} b_{0\mu} b_0^{\mu}\nonumber  \\
 && - \frac{1}{4} A_{\mu \nu}A^{\mu \nu}+\frac{1}{2}(\partial _{\mu}\sigma \partial ^{\mu} \sigma -
  m^2_{\sigma}\sigma^2) - \frac{1}{3}g_2\sigma^3 - \frac{1}{4}g_3\sigma^4 + \mathcal{L}_K,
\end{eqnarray}
where $\psi_B$, $\sigma$, $\omega_{\mu}$, and ${b}_{0\mu}$ are the
fields of the baryon, scalar, vector, and charge-neutral
isovector-vector mesons, with their masses $M_B$, $m_{\sigma}$,
$m_{\omega}$, $m_{\rho}$, respectively. The $A_{\mu}$ is the field of
photon. The $g_{iB} (i=\sigma, \omega, \rho)$ are the corresponding
meson-baryon coupling constants. The $\tau_3$ is the third component of
isospin Pauli matrix for nucleons. $F_{\mu \nu}$, $B_{\mu \nu}$, and
$A_{\mu \nu}$ are the strength tensors of the $\omega$, $\rho$ meson
and the photon, respectively
$$F_{\mu
\nu}=\partial_{\mu}\omega_{\nu}-\partial_{\nu}\omega_{\mu},\hbox{ }
B_{\mu \nu}=\partial_{\mu}b_{0\nu}-\partial_{\nu}b_{0\mu}, \hbox{ }
A_{\mu \nu}=\partial_{\mu}A_{\nu}-\partial_{\nu}A_{\mu}.$$
 $\mathcal{L}_K$, the lagrangian of the  kaonic sector~\cite{07DG,08DG}, is written as
\begin{equation} \label{lagK}
\mathcal{L}_K = (\mathcal{D}_{\mu}K)^{\dag}(\mathcal{D}^{\mu}K)
 - (m^2_{K} - g_{\sigma K} m_{K} \sigma )K^{\dag}K,
\end{equation}
where the covariant derivative is given by
\begin{equation} \label{der}
 \mathcal{D}_{\mu} \equiv \partial _{\mu} + i g_{\omega K} \omega_{\mu} + i g_{\rho K}
 b_{0\mu} + ie\frac{1+\tau_3}{2}A_{\mu},
\end{equation}
with the $g_{iK} (i=\sigma, \omega, \rho)$ being the corresponding
$K^-$-meson coupling constants. Here,  $K$ and $K^\dag$ denote the
kaon and antikaon doublet, i.e. $K=\binom{K^+}{K^0}$ and $K^\dag
=(K^-,\bar{K}^0)$, respectively. Since our investigation is limited
to $K^-$ nuclei, the $K^0$ and $\bar{K}^0$ are left out in the
following.

The equations of motion are derived from the Lagrangian~(\ref{Lag}). In the RMF
approximation, the nucleons obey the following equation
\begin{equation}
[-i\vec{\alpha}\cdot \vec{\nabla} + \beta (M_B -g_{\sigma
B}\sigma_0)+ g_{\omega B}\omega_0 + g_{\rho B}\tau_3
b_0+e\frac{1+\tau_3}{2}A_0]\psi_B=E_B\psi_B. \label{EOMd}
\end{equation}
In spherical systems we consider in this work, the equations of
motion for non-strange mesons are given by
\begin{eqnarray}
&& (\frac{d^2}{dr^2} + \frac{2}{r} \frac{d}{dr} -
m^2_{\sigma})\sigma_0 = g_{\sigma B}\rho_s - g_2\sigma^2_0 -
g_3\sigma^3_0 + g_{\sigma K}m_K K^-K^+,\nonumber
\label{EOMs}\\
&&(\frac{d^2}{dr^2}+\frac{2}{r} \frac{d}{dr}-m^2_{\omega})\omega_0
=g_{\omega B}\rho_v-g_{\omega K}\rho_{K^-},\nonumber
\label{EOMo}\\
&&(\frac{d^2}{dr^2}+\frac{2}{r} \frac{d}{dr}-m^2_{\rho})b_0 =g_{\rho
B}\rho_3-g_{\rho K}\rho_{K^-},
\label{EOMr}\\
&&(\frac{d^2}{dr^2}+\frac{2}{r}\frac{d}{dr})A_0=e\rho_p -
e\rho_{K^-}, \label{EOMe}\nonumber
\end{eqnarray}
where $\rho_s$, $\rho_v$, $\rho_p$ and $\rho_3$ are the scalar,
vector, proton, and isovector densities, respectively, and  we refer
readers to Ref.~\cite{86Adv, 05Jiang} for detailed expressions. The
$\rho_{K^-}$ denotes the $K^-$ density
\begin{equation}
\rho_{K^-}=2(E_{K^-}+g_{\omega K}\omega_0+g_{\rho K} b_0+eA_0)K^-K^+,
\end{equation}
where the integration of $\rho_{K^-}$ over the whole volume is
normalized to the $K^-$ number which is one in this work. The
Klein-Gordon equation for $K^-$ reads
\begin{equation}\label{KG-K}
(\nabla^2+E^2_{K^-}-m^2_K-\Pi)K^-= 0,
\end{equation}
where $E_{K^-}$ is the single-particle energy of the $K^-$. The real
part of the $K^-$ self-energy $\Pi$ is written as
\begin{eqnarray}
\Re \Pi&=&-g_{\sigma K}m_K\sigma_0 - 2E_{K^-}(g_{\omega
K}\omega_0+g_{\rho K} b_0+eA_0)\nonumber\\
 &&- (g_{\omega K}\omega_0+g_{\rho K} b_0+eA_0)^2.
\end{eqnarray}
The imaginary part $\Im\Pi$ is considered as the absorptive
contribution  to the $K^-$ self-energy which can be taken from some optical models
phenomenologically. Following works of Mare\v{s} et al. ~\cite{05JM,
06JM}, we adopt the simple '$t\rho$' form $\Im\Pi=fV_{0}\rho_{v}(r)$
where the depth $V_0$ is obtained by fitting $K^-$ atomic
data~\cite{99EF}.

\section{Revisit to the relativistic symmetry manifestation in RMF}
\label{PSS}

The concept of PSS is introduced to describe the quasidegeneracy in
some nuclei between single-nucleon states with quantum numbers $(n, l
, j = l + 1/2)$ and $(n-1, l+2, j = l+ 3/2)$, e.g., $2S_{1/2}$ and
$1D_{3/2}$. The PSS can be understood as a relativistic symmetry of
the Dirac Hamiltonian originating from the near equality in magnitude
of the scalar potential S(r) and vector potential V(r) but different
in sign, i.e. $S(r)+V(r) \approx 0$ ~\cite{97JNG}. The quality of the
approximate symmetry was found to be associated with the competition
between pseudocentrifugal barrier and the pseudospin-orbit
potential~\cite{98JM,99JM}. In the following, we demonstrate in the
RMF the conditions for the PSS.

Since the meson fields are classical in the RMF, one can write the
Dirac Hamiltonian for  nucleons in spherical nuclei as
\begin{equation}
\emph{\^{h}} = -i\vec{\alpha} \cdot \vec{\nabla} +
 g_{\omega B}\omega_0+g_{\rho B}\tau_3 b_0+e\frac{1+\tau_3}{2}A_0+\beta
 (M_B - g_{\sigma B}\sigma_0) .
\end{equation}
It is well-known that the total  angular momentum \textbf{J}
commutates with the Hamiltonian and is a conserved operator. While
the orbital angular momentum  $\bf{L}$ does not commute with the
\emph{\^{h}}, to characterize the full set of quantum numbers of
single-particle states, one invokes another conserved operator
\textbf{K}
\begin{equation}
\mathbf{K}\equiv -\gamma^0 (\mathbf{\Sigma \cdot L}+1),
\end{equation}
with its eigenvalues
\begin{equation}
\kappa= \left\{
\begin{array}{cl}
\emph{l},\quad &  \emph{j=l}-\frac{1}{2},\\
-(\emph{l}+1),\quad & \emph{j=l}+\frac{1}{2}.
\end{array}
\right .
\end{equation}
Then we have $j=|\kappa |-1/2 $ and
$l=|\kappa|+(\kappa/|\kappa|-1)/2$. For a spherical system, the
quantum number set is $\{n,\kappa,m, t\}$ where \emph{n} is the
principle quantum number, \emph{m} is the magnetic quantum number, and $t$ denotes the isospin. With these
quantum numbers, the single-particle wave functions can be written as
\begin{equation}\label{GF}
\psi_{n\kappa m t}(r)=\left\{\begin{array}{c} i\frac{G_a(r)}{r}\Phi_{\kappa,m}\\
-\frac{F_a(r)}{r}\Phi_{-\kappa,m}\end{array} \right\} \chi_t,
\end{equation}
where $G_a(r)$ and $F_a(r)$ are the big and small components of the
(radial) spinor, respectively, $\{a\}=\{n,\kappa,t\}$, $\Phi$ is the spinor spherical harmonic, and
$\chi_t$ is the isospinor with $t=\pm1$ for protons and neutrons,
respectively. Substituting Eq. (\ref{GF}) into the Dirac equation
(\ref{EOMd}),  one can immediately obtain the radial equations for
nucleons
\begin{eqnarray}
(\frac{d}{dr}+\frac{\kappa}{r})G_a(r) &=& (M_B + E_a - \triangle)F_a(r)\label{up},\\
(\frac{d}{dr}-\frac{\kappa}{r})F_a(r) &=& (M_B - E_a +
\Sigma)G_a(r)\label{low},
\end{eqnarray}
where
\begin{equation}
\Sigma = V(r)+S(r), \hbox{ }\triangle=V(r)-S(r),
\end{equation}
with $V(r)=g_{\omega B}\omega_0+g_{\rho B}\tau_3
b_0+eA_0(1+\tau_3)/2$ and $S(r)=-g_{\sigma B}\sigma_0$. By performing
radial derivative on the both sides of  Eq.~(\ref{up})
and~(\ref{low}), one arrives at the two second-order differential
equations for the big and small components, respectively
\begin{eqnarray}
\label{eqG}
[\frac{d^2}{dr^2}+\frac{1}{U_G}\frac{d\triangle}{dr}\frac{d}{dr}
+\frac{1}{U_G}\frac{d\triangle}{dr}\frac{\kappa}{r}-\frac{\kappa(\kappa+1)}{r^2}-U_GU_F
]G_a(r) &=& 0,\\
\label{eqF}
[\frac{d^2}{dr^2}-\frac{1}{U_F}\frac{d\Sigma}{dr}\frac{d}{dr}
+\frac{1}{U_F}\frac{d\Sigma}{dr}\frac{\kappa}{r}-\frac{\kappa(\kappa-1)}{r^2}-U_GU_F
]F_a(r) &=& 0,
\end{eqnarray}
with $U_G=M_B+E_a-\triangle$ and $U_F=M_B-E_a+\Sigma$.  These two
equations are equivalent for obtaining the eigenvalues $E_a$. Similar
to the description in Ref.~\cite{98JM}, the centrifugal barrier (CB)
and pseudocentrifugal barrier (PCB) are here defined as
$\kappa(\kappa+1)/r^2$ and $\kappa(\kappa-1)/r^2$, respectively. The
spin-orbit potential (SOP) and pseudospin-orbit potential (PSOP)
terms are those in Eqs.(\ref{eqG}) and (\ref{eqF}) proportional to
$\kappa d\triangle/(rdr)$ and $\kappa d\Sigma/(rdr)$, respectively.
If the SOP term equals exactly to zero, say, $\triangle=0$ or
$d\triangle/dr=0$, it can be obviously seen from Eq.~(\ref{eqG}) that
the eigenvalue $E_a$ only depends on $\kappa(\kappa+1)$. One can
easily find out that those states with $\kappa=l$ and $\kappa=-(l+1)$
share the same eigenvalues. This is nothing but the spin symmetry
leading to the spin degeneracy. Similarly, if the PSOP term equals
exactly to zero, say, $\Sigma=0$ or $d\Sigma/dr=0$, then the
eigenvalue $E_a$ only depends on $\kappa(\kappa-1)$ as seen from
Eq. (\ref{eqF}). Introducing $\tilde{l}= l+1$, we can see that those
states with  $\kappa=-\tilde{l}$ and $\kappa=\tilde{l}+1$ have the
same eigenvalues. This type of degeneracy was named the PSS in
resemblance to the spin symmetry. However, the PSS is nonexistent
once the PSOP term is far away from zero. Generally speaking, as
already pointed out in Ref.~\cite{98JM}, the quality of the PSS is
tightly associated with the relative magnitude of PCB to the PSOP in
the RMF, while the correlation between PSOP term and the PSS is not
simply linear~\cite{02PA}.

In a normal nuclear system, the attractive potential is usually
around 380MeV, while the repulsion is around 320MeV. It gives rise to
$\triangle\approx700$MeV, corresponding to relative large SOP term,
while the $\Sigma$ is around $-60$ MeV, leading to a relative small
PSOP term. This estimation indicates that the PSS is
developed  much  better than the spin symmetry in normal nuclear
single-particle spectra. In plenty of nuclei, the PSS
exists approximately and the splitting of pseudospin doublets $(n, l
, j = l + 1/2)$ and $(n-1, l+2, j = l+ 3/2)$ is relatively small,
compared to the separation between two levels nearby. But the
appearance of $K^-$ could change the proportion of the attraction to
the repulsion, leading to the alteration of $\triangle$ and $\Sigma$.
Thus, the general situation for the PSS in kaonic
nuclei can be rather different and it is worthy of careful
investigations.

\section{Results and discussions}
\label{result} As one source term of the mean field, the $K^-$
diminishes its role in the mean field with the increase of the
nuclear mass. Thus, our investigation is limited to the medium and
light nuclei in which the $K^-$ has more distinct effects on nuclear
properties, especially the bulk density of the nuclear system and the
single-particle energy for the pseudospin and spin doublets. We
perform calculations with the NL3 parameter set~\cite{97GAL}. The
$g_{\omega K}$ and $g_ { \rho K}$ are chosen from SU(3) relations:
$2g_{\omega K}=2g_ { \rho K}=g_ { \rho \pi}=6.04$, and the $g_{\sigma K}$
is adjusted to yield a $K^-$ binding energy $B_{K^-}$=100MeV for $^{40}_{K^-}$Ca where the $B_{K^-}$ is defined
as the difference of the total binding energy between the kaonic
nucleus and its normal counterpart without $K^-$. Though the
moderately deep optical potential is adopted here ($\approx 100$ MeV), as
in some pioneer works~\cite{06XHZ,07DG,13XRZ}, we will examine the
cases for various binding energies at last. The coupled equations
(\ref{EOMr}),(\ref{KG-K}), (\ref{up}) and (\ref{low}) are solved
self-consistently by an iterative procedure. Some details of solving
the $K^-$ equation are given in the Appendix.

\begin{table}[htp]
\caption{ Single-neutron binding energies and splittings of the
pseudospin and spin doublets in normal nuclei and the corresponding
kaonic nuclei (in unit of MeV) with the NL3. \label{spectra}}
\begin{center}
\begin{tabular}{ c | c   c   c || c  c  |c  c  c }
\hline\hline
 & $2S_{1/2}$ & $1D_{3/2}$ & $\triangle$ (2S-1D) & $1D_{5/2}$ & $\triangle$ (1D)& $1P_{3/2}$ & $1P_{1/2}$  & $\triangle$ (1P) \\
\hline
 $^{16}$O & - & - & - &- & - &  21.73 & 15.25 & 6.48 \\
$_{K^-}^{16}$O & - & - & - &- & - &  27.43 & 6.75 & 20.68\\
\\
 $^{34}$S & 13.95 & 10.45 & 3.50 & 18.59 & 8.14 & 35.85 & 28.82 & 7.03 \\
$_{K^-}^{34}$S & 20.81 &8.45 &12.36 &18.85 &10.40& 40.18& 24.52& 15.66\\
\\
 $^{40}$Ca & 16.96 & 16.17 & 0.79 & 22.88 & 6.71 & 37.98 & 33.50 & 4.48
\\
$_{K^-}^{40}$Ca &25.86&14.49&11.37&23.19& 8.70&41.57& 30.94&10.63\\
\\
$^{48}$Ca & 17.56 &17.73&-0.17&23.88&6.15 &38.94 &35.63& 3.31\\
$_{K^-}^{48}$Ca & 21.36 & 16.77&4.59& 25.00& 8.23 &42.61 &36.48&6.13\\
\\
$^{52}$Cr & 20.21 &21.89 &-1.68 &27.95 &6.06 &42.95 &40.11& 2.84\\
$_{K^-}^{52}$Cr & 22.95 & 21.44 &1.51 & 29.52 & 8.08 & 46.80 & 42.12&4.68\\
\\
$^{58}$Ni & 22.85 & 26.08 & -3.23 & 31.38 & 5.30 & 45.71 & 43.67&2.04 \\
$_{K^-}^{58}$Ni & 25.10 & 26.22 & -1.12 & 33.23&7.01&49.59&46.41&3.18\\
\\

$^{74}$Se & 25.63 & 29.42 & -3.79 & 33.55 & 4.13 & 45.21 & 43.41 & 1.80 \\
$_{K^-}^{74}$Se &28.05&29.78&-1.73& 34.90 & 5.12 & 47.93 & 45.40 & 2.53 \\
\\

$^{90}$Zr & 30.29 & 32.43 & -2.14 & 36.32 & 3.89 & 48.38 & 46.56 & 1.82 \\
$_{K^-}^{90}$Zr & 33.35& 32.62 &0.73& 37.46 & 4.84 & 51.02 & 48.14 & 2.88 \\
\hline
\hline
     \end{tabular}
     \end{center}
  \end{table}

Tabulated in Table~\ref{spectra} are the single-neutron binding
energies and the splittings of pseudospin and spin doublets for a few
states in some spherical  nuclei and in the corresponding kaonic
nuclei. Those results for protons are not presented here because of
no qualitative difference therein. Seen from this table, the most
noticeable change of the single-particle binding energy in these
kaonic nuclei is the considerable increase in the binding energy of
the $2S_{1/2}$ and $1P_{3/2}$ states while just a little effect on
$1D_{3/2}$, especially for those light nuclei. For instance, in
$^{40}$Ca, the pseudospin doublets $2S_{1/2}$ and $1D_{3/2}$ are very
close,  showing that the PSS is satisfied approximately, whereas the
situation is quite different when the $K^-$ is implanted into
$^{40}$Ca. The big separation between the corresponding doublets
arises in $_{K^-}^{40}$Ca due to a dramatic increase of the binding
energy of the $2S_{1/2}$ state while a small reduction in $1D_{3/2}$,
leading to the manifest breaking of the PSS. Besides in
$_{K^-}^{40}$Ca, similar orbital shifts take place in other kaonic
nuclei as seen in Table~\ref{spectra}.  Interestingly, with the
increase of the nucleon number, the orbital shifts in heavier nuclei
$_{K^-}^{58}$Ni, $_{K^-}^{74}$Se, and $_{K^-}^{90}$Zr even favor the
approximate PSS, being more satisfactory than those in corresponding
normal nuclei. This takes place   because the shallower binding of
the $2S_{1/2}$ state in the pseudospin doublet of heavier nuclei
accompanies with the smaller binding enhancement of such a state in
corresponding kaonic nuclei. On the other hand, the spin-orbit
splittings between $1P_{3/2}$ and $1P_{1/2}$ all get a rise in kaonic
nuclei listed in Table~\ref{spectra}. For instance, the energy
interval between $1P_{3/2}$ and $1P_{1/2}$ is 4.48 MeV for $^{40}$Ca,
while it is 10.63 MeV for $_{K^-}^{40}$Ca. Needless to say, the
breaking of the spin symmetry becomes more distinctive in these
kaonic nuclei, in comparison to that in normal nuclei. We mention
that the effect of the imaginary part of $K^-$  on the nuclear
density and single-particle energies is insignificant for the given
$K^-$ binding energy 100 MeV, and it just becomes moderate for much
weaker binding energies. Thus, we will not regard it specifically in
the following discussion.

\begin{figure}[thb]
\centering
\includegraphics[height=7.0cm,width=8.6cm]{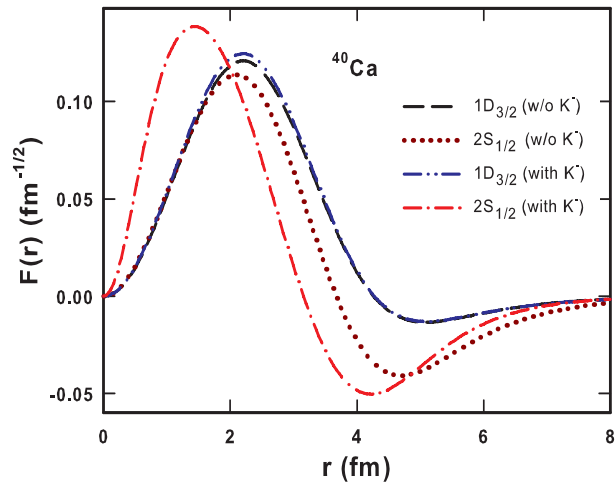}
\caption{(Color online) The small components of the Dirac spinor for
neutrons in $^{40}$Ca and $_{K^-}^{40}$Ca as a function of radius.
Those labelled "w/o $K^-$" and "with $K^-$" represent for normal
nuclei and corresponding kaonic nuclei, respectively, and we will
keep these abbreviations throughout.} \label{WF}
\end{figure}

Now, we take $^{40}$Ca and $_{K^-}^{40}$Ca as a typical example to
understand the PSS features in kaonic nuclei.  In Fig.~\ref{WF}, we
plot  the small  components of the Dirac spinor, i.e. $F_a(r)$. As
discussed in Sec.\ref{PSS}, the similarity in the small components of
the Dirac spinor is associated with the approximate
PSS~\cite{98JNG}. Indeed, this is the case in $^{40}$Ca: the $F(r)$s
for the pseudospin doublets $2S_{1/2}$ (dotted curve) and $1D_{3/2}$
(dashed curve) are rather close, though one has a node and the other
without. But the addition of the $K^-$ in $^{40}$Ca forces the small
component of the $2S_{1/2}$ state to move inwards considerably. As a
result, the breaking of the PSS appears with the similarly explicit
shift in the small component of the Dirac spinor. While with the
increase of the nuclear number, the relative shift of $F(r)$ in
kaonic nuclei becomes  trivial, consistent with the much smaller
shift in single-particle binding energy.

\begin{figure}[thb]
\centering
\includegraphics[height=10.0cm,width=8.6cm]{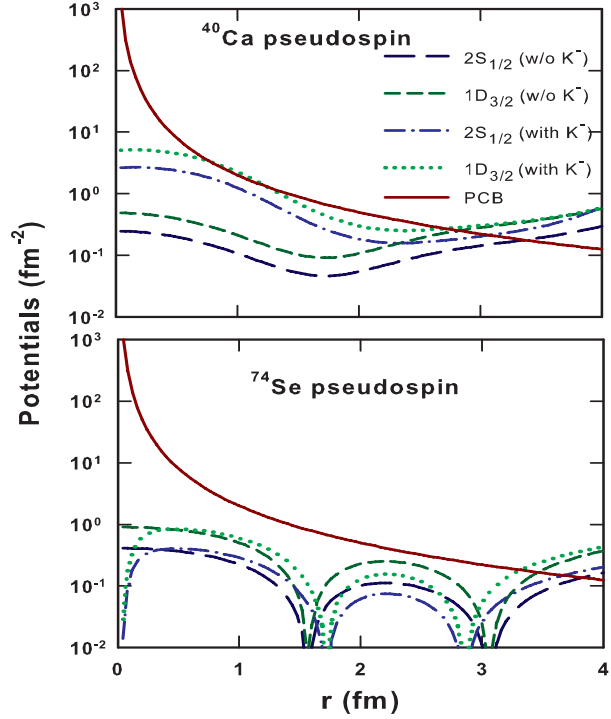}
\caption{(Color online) The PCBs and PSOPs (denoted by orbital quanta)
in Eq.~(\ref{eqF}) for pseudospin doublets $2S_{1/2}$ and $1D_{3/2}$
in $^{40}$Ca and $^{74}$Se systems. Also see text. } \label{POTps}
\end{figure}

\begin{figure}[thb]
\centering
\includegraphics[height=10.0cm,width=8.6cm]{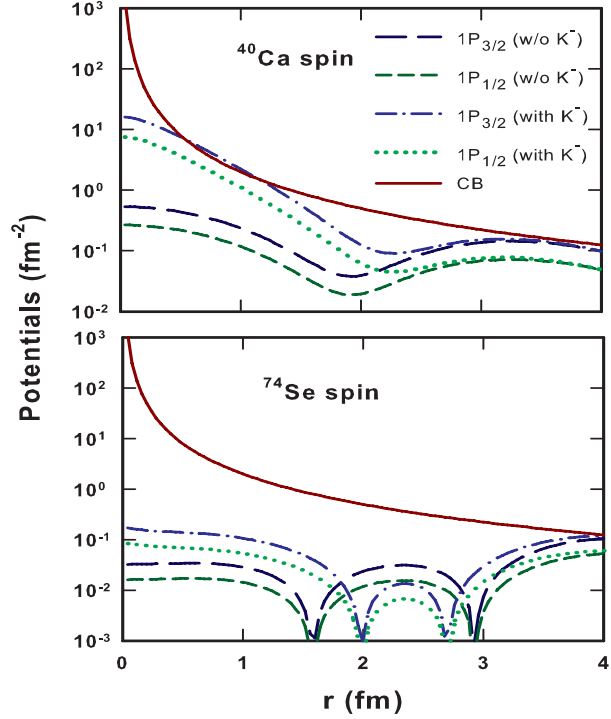}
\caption{(Color online) The CBs $|\kappa(\kappa-1)|/r^2$  and SOPs
 $|\kappa d\Sigma/U_F r dr|$ (labelled by orbital quanta), see
Eq. (\ref{eqG}), for spin doublets $1P_{1/2}$ and $1P_{3/2}$ in
$^{40}$Ca and $^{74}$Se systems.} \label{POTs}
\end{figure}

For a deeper understanding of these phenomena concerning the addition
of  the $K^-$, we illustrate in Fig.~\ref{POTps} the radial
distributions of some relevant potential terms for pseudospin
doublets ($2S_{1/2}$ and $1D_{3/2}$) in $^{40}$Ca, $^{74}$Se and in
the corresponding kaonic nuclei. They are the PCB
$|\kappa(\kappa-1)|/r^2$ and the PSOP $|\kappa d\Sigma/U_F r dr|$,
see Eq. (\ref{eqF}). Because the $K^-$ is trapped in the interior
nutshell by the strong attraction and its effect becomes
insignificant beyond $r=4fm$, only shown in the figure are the
results in the region $r\leq 4fm$ for clarity. The relative magnitude
of the PCB to the PSOP, closely associated with the quality of the
PSS,  can be used to evaluate the extent of the corresponding
symmetry breaking. It is seen from the upper panel of
Fig.~\ref{POTps} that PSOP is largely enhanced by the $K^-$ in
$_{K^-}^{40}$Ca, especially in the region $r\lesssim 2fm$. Due to the
large enhancement, the PSOP and PCB in $_{K^-}^{40}$Ca are
comparable. It is now not difficult for us to understand why  the PSS
is destroyed in $_{K^-}^{40}$Ca:  the clear enhancement of the PSOP
brings out the importance of the $\kappa$ dependence of eigen
energies and vectors (wave functions) that is the exact factor for
the PSS breaking, see Eq. (\ref{eqF}). We have already seen in
Table~\ref{spectra} that the PSS in $_{K^-}^{74}$Se is a little
better than that in $^{74}$Se. It is again illustrated in terms of
the corresponding potentials for $^{74}$Se and $_{K^-}^{74}$Se in the
lower panel of Fig.~\ref{POTps} that the PSOP is small compared to
the PCB and is much less affected by the $K^-$, which justifies the
approximate PSS both in $^{74}$Se and $_{K^-}^{74}$Se. We may further
examine the potentials in $^{74}$Se and $_{K^-}^{74}$Se in more
details.  For $r\lesssim 1.6fm$, the PSOP difference between
$^{74}$Se and $_{K^-}^{74}$Se is not significant, while in the range
of $1.6fm\lesssim r \lesssim 2.8fm$, the PSOP of $^{74}$Se is clearly
larger than that of $_{K^-}^{74}$Se. The latter may be responsible
for the improvement of the PSS in $_{K^-}^{74}$Se, as this is
consistent with the situation in heavier nuclei that more nucleons in
exterior shells produce the more radial extension of the $K^-$ by the
attraction. In general, the difference in the PSOP caused by the
$K^-$ embedment becomes small in heavy nuclei, and distinctive
features with the embedment of the $K^-$ tend to disappear.

The situation for the shift of spin-orbit splitting is quite
analogous. The attraction provided by the $K^-$ leads to the
deepening of the SOP, and the $\kappa$ dependence
of eigenvalues of the spin doublets is magnified. For heavier and
heavier nuclei, the $K^-$ effect on the spin-orbit splitting drops
off, just like the case of the PSS breaking. These
phenomena are clearly shown in Fig.~\ref{POTs}, in resemblance to the
pseudospin case in Fig.~\ref{POTps}, and are consistent with those
for the spin-orbit splittings listed in Table~\ref{spectra}.

\begin{figure}[htb]
\centering
\includegraphics[height=10.0cm,width=15.0cm]{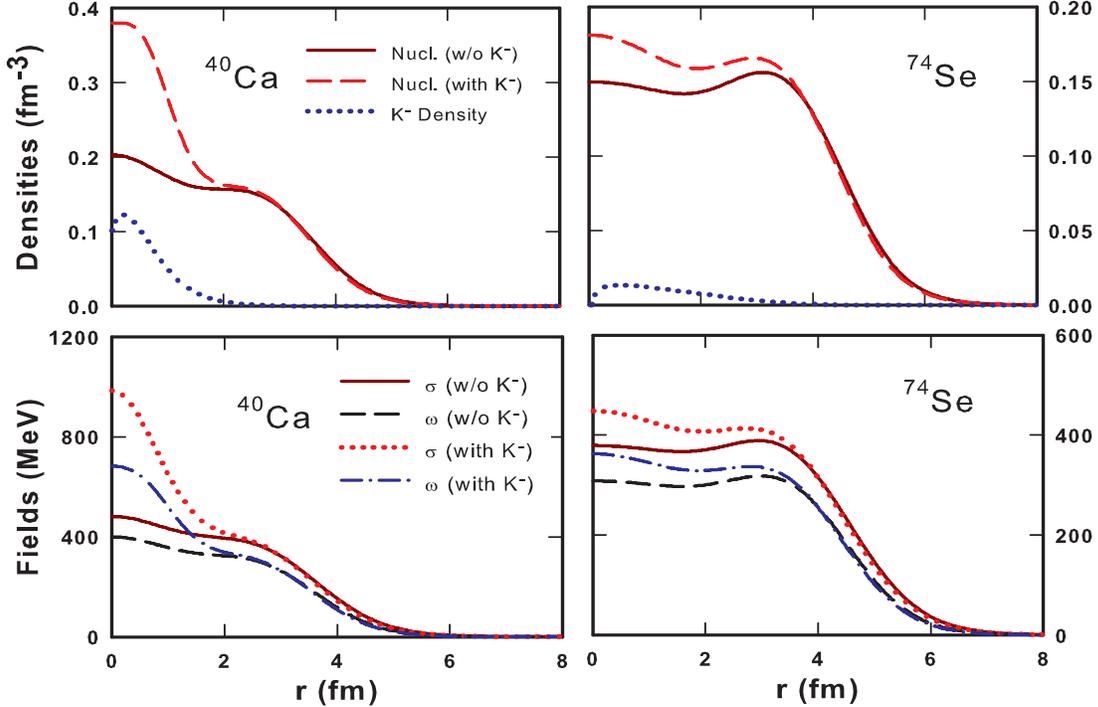}
\caption{(Color online)Nuclear densities and potentials as a function
of radius in $^{40}$Ca (left panels) and $^{74}$Se systems (right
panels). } \label{dens}
\end{figure}

The most striking phenomenon of the kaonic nuclei should be the
enhancement of the central density~\cite{06XHZ,07DG,13XRZ}. The
enhancement in light nuclei could be very prominent. Here, we  take
$^{40}$Ca and comparably $^{74}$Se systems as examples. In
Fig.~\ref{dens}, it displays the radial distributions of the
densities and potentials. For $^{40}$Ca systems, we see that  the
$K^-$ is deeply bound inside the interior of the kaonic
nuclei($r\lesssim 2fm$) due to the strong attraction. This deeply
trapped $K^-$ distribution is consistent with the sizeable rise of
the SOP and PSOP in the core region, as
shown in Fig.~\ref{POTps} and~\ref{POTs}, respectively. As a result,
the strong attraction provided by the $K^-$ in $^{40}$Ca pulls
nucleons inwards to form a dense core with a density  up to twice the
saturation density, as shown in the upper left panel of
Fig.~\ref{dens}. We recall that the strong attraction responsible for
this is produced by the coherent coupling of $\sigma$ and $\omega$
mesons to the $K^-$. Feeding back to the mean field in kaonic nuclei,
the attractive potential ($g_{\sigma}\sigma$) in the shrunk core
acquires an enhancement greater than the one for the repulsive
potential ($g_{\omega}\omega$), as clearly shown in the lower left
panel in Fig.~\ref{dens}. The enhancement of the core density also
exists in all the kaonic nuclei, while it fades away gradually in
heavy nuclei. Shown in the right panels of Fig.~\ref{dens} is the
case for $^{74}$Se systems. As can be seen, the $K^-$ is pulled
outwards by the attraction provided by more out-layer nucleons. As a
consequence, just moderate increase of the core density and
mean-field potentials in $_{K^-}^{74}$Se, instead of the dramatic
increase in light nuclei, is observed. Consistently, the shifts of
single-particle energy are much smaller than those in light nuclei.
We note that the calculations with the non-relativistic
Skyrme-Hartree-Fock approach also found the similar
shrinkage~\cite{13XRZ}. These results are inspiring as kaonic nuclei
could probably provide a natural cold dense nuclear system rather
than a hot dense one that should be created by heavy-ion collisions.

\begin{figure}[htb]
\centering
\includegraphics[height=6.0cm,width=8.6cm]{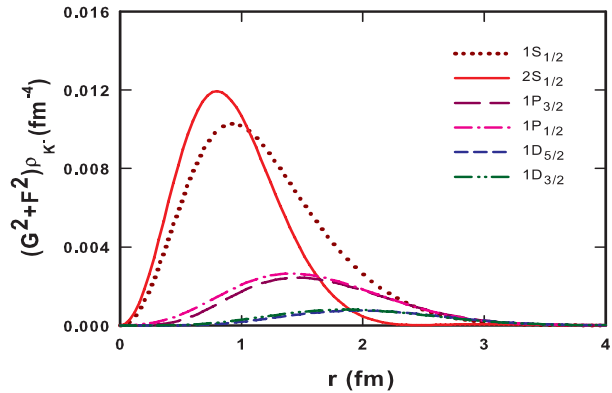}
\caption{(Color online) Products of the square of  various neutron
wave functions  in $^{40}$Ca and $K^-$  density distribution in
$_{K^-}^{40}$Ca. } \label{overlap}
\end{figure}

\begin{figure}[thb]
\centering
\includegraphics[height=7.0cm,width=8.6cm]{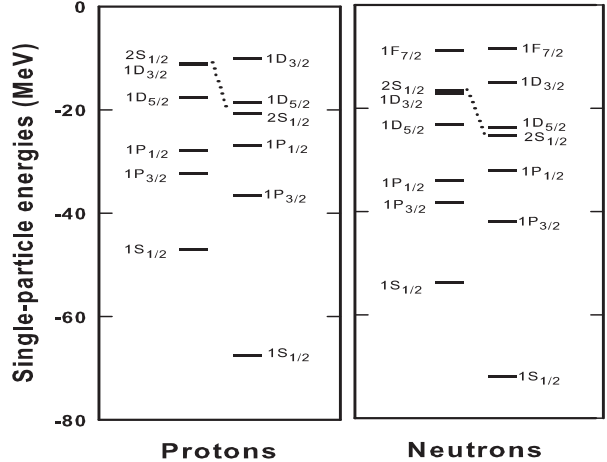}
\caption{ Single-particle energies in $^{42}$Ca and $^{42}_{K^-}$Ca.
The left column  in each panel represents energy levels for
$^{42}$Ca, while the right ones are for $^{42}_{K^-}$Ca}. \label{spe}
\end{figure}

Besides the above general analysis for the PSS breaking, we have not
paid much attention to a specific phenomenon, the almost unilateral
shift of the pseudospin doublets. Looking back to Fig.~\ref{WF},  we
see that it is the shift of the 2$S_{1/2}$, rather than that of the
1$D_{3/2}$, which dominates the PSS breaking.  This can be roughly
understood by the spatial proximity between the $1S_{1/2}$ $K^-$ and
$2S_{1/2}$ nucleons. To make it clear,  we plot the simple product
between the neutron wave function square of different states in
$^{40}$Ca and $K^-$ density distribution in $_{K^-}^{40}$Ca in
Fig.~\ref{overlap}. We see that unlike the $S_{1/2}$ orbitals, the
$1D$ orbitals are almost uncorrelated with the $K^-$ occupation. This
illustrates why the $2S_{1/2}$ state of pseudospin doublets is
affected much more dramatically by the $K^-$. Further, this is
associated with the change of the shell structure. Shown in
Fig.~\ref{spe} is the single-particle spectra for protons and
neutrons in $^{42}$Ca and $^{42}_{K^-}$Ca. It is clearly shown that
the $K^-$ implantation leads the migration of $1S_{1/2}$ and
$2S_{1/2}$ nucleons downwards to the deep Fermi sea. The orbital
migration certainly changes the original magic numbers in normal
nuclei that are 2, 8, 20... In $^{42}_{K^-}$Ca, the magic numbers now
become 2, 6 and 16 that are dictated  by the migrated $S_{1/2}$
orbitals and enlarged spin-orbit splittings of $1P$ and $1D$
orbitals, as shown in Fig.~\ref{spe}. Though it is perhaps premature
to speak of the new magic numbers because the magic gaps should rely
on the depth of the potential well for the $K^-$, it is definite that
the nuclear structure is changed by the $K^-$ implantation.

\begin{figure}[thb]
\centering
\includegraphics[height=6.0cm,width=8.6cm]{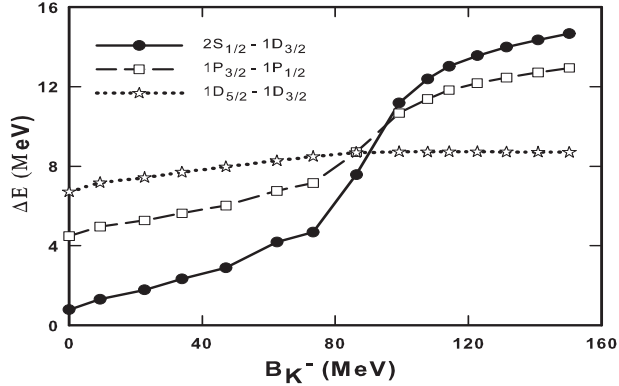}
\caption{ Splittings of pseudospin and spin doublets in $_{K^-}^{40}$Ca
as  a function of $K^-$ binding energy. $B_{K^-}$=0 represents normal
$^{40}$Ca. \label{LS-B} }
\end{figure}

By now, there is no conclusive value of the depth of the
$K^-$-nuclear potential. Thus, it is valuable to investigate the
dependence of energy splittings of the pseudospin and spin doublets
on the $K^-$ binding energy. Specifically, we obtain various
splittings of pseudospin and spin doublets  by changing
$g_{{\sigma}K}$ and $g_{{\omega}K}$, which is equivalent to varying
the $K^-$ binding energy.  Results are displayed in Fig.~\ref{LS-B}
for the pseudospin doublets ($2S_{1/2}$ and $1D_{3/2}$), ($1P_{1/2}$
and $1P_{3/2}$) and ($1D_{3/2}$ and $1D_{5/2}$). At the point
$B_{K^-}$=0, i.e. the normal $^{40}$Ca, the pseudospin doublet
splitting between $2S_{1/2}$ and $1D_{3/2}$ is less than 1 MeV. The
splitting becomes increasingly large with the enhancement of the
$K^-$-nuclear attraction, while a clear increase appears at
$B_{K^-}>80MeV$. Similarly, the splitting of the spin doublets
$1P_{1/2}$ and $1P_{3/2}$ is apparently amplified by increasing the
$K^-$-nuclear attraction. We see that the effect on the doublet
$1D_{3/2}$ and $1D_{5/2}$ is relatively smaller. This is
understandable since the out-layer states are less affected by the
interior $K^-$. Nevertheless, larger $K^-$ binding can generally
result in more prominent phenomenon of the pseudospin and spin
symmetry breakings, especially for interior states.

The above discussion on the association between the doublet
splittings and the $K^-$ binding energy also leads our attention to
more details of the model dependence. We find that the splittings of
the pseudospin and spin doublets are almost independent of various
combinations of different $g_{{\sigma}K}$ and $g_{{\omega}K}$ for a
given $K^-$ binding energy. While we change to other models like the
NL-SH~\cite{MMS93},  we find that the size of the doublet splittings
is quantitatively different not only due to various incompressibility
at saturation density but also because of the rather different
equations of state at suprasaturation densities. Nevertheless, the
splittings are still very large for light kaonic nuclei. Moreover, we
should point out that all results with the NL-SH are qualitatively
similar to those obtained from the NL3, without changing the
conclusions drawn above.

At last, it is worthy to point out the dependence of the spin or
pseudospin doublet splittings on the isospin. With the increase of
the neutron number, the repulsion provided by the isovector meson
increases. This causes the moderate reduction of the relevant doublet
splittings. However, the splittings are still well above those for
the normal nuclei.

\section{Summary}
\label{summary} We have investigated the ground-state properties of
light kaonic nuclei and the relativistic symmetry breakings with the
RMF model.  It is found that  the relativistic symmetry breakings,
underlying potentials associated with the breakings, and the
appreciable shrinkage effect, i.e., the significantly enhanced core
density take place consistently due to the strong attraction provided
by the $K^-$ embedment in light nuclei. For normal nuclei
entertaining the approximate pseudospin symmetry, the $K^-$ embedment
can enhance $\kappa$-dependence of the eigen energies and wave
functions, break the original balance between the attractive and
repulsive potentials consequently, and result in  the pseudospin
symmetry breaking and the spin symmetry deterioration.  With the
$K^-$ embedment, the shell structures are also changed dramatically
in light nuclei. In particular, the migration of $S_{1/2}$ orbitals
towards the interior Fermi sea and the enlarged spin-orbit splittings
can reshape the order of energy levels and form new magic numbers.
Moreover, we have examined the dependence on the $K^-$ binding energy
and the model dependence for these phenomena. The model dependence is
rather weak, and we find that for a large parameter space of the
$K^-$ binding energy all the phenomena are quite general.

\section*{Acknowledgements}
We thank Prof. Xian-Hui Zhong for useful discussions. The work was
supported in part by the National Natural Science Foundation of China
under Grant No. 11275048,  the China Jiangsu Provincial Natural
Science Foundation under Grant No.BK20131286, and the China
Scholarship Council.
\appendix
\section{Solving the Klein-Gordon equation for $K^-$}
\label{append} We give some necessary  details for solving the
Klein-Gordon equation for $K^-$ (Eq.~\ref{KG-K}) herein.
Eq. (\ref{KG-K}) is a complex equation which should firstly be divided
into two coupled real equations. In a spherical system,
Eq. (\ref{KG-K}) can be written as
\begin{equation}
(-\frac{d^2}{dr^2}+\frac{l(l+1)}{r^2}+M_K^2+\Pi(r))K^-(r)=E^2K^-(r).\label{eqK}
\end{equation}
It is decomposed as
\begin{eqnarray}
(\frac{d^2}{dr^2}+U_1(r))K^-_{\Re}(r)-U_2(r)K^-_{\Im}(r)=0, \label{ReK}\\
(\frac{d^2}{dr^2}+U_1(r))K^-_{\Im}(r)+U_2(r)K^-_{\Re}(r)=0,
\label{ImK}
\end{eqnarray}
with $U_1(r)=-\frac{l(l+1)}{r^2}+E_{\Re}^2-E_{\Im}^2-M_K^2-\Re
\Pi(r)$, $U_2(r)=2E_{\Re}E_{\Im}-\Im \Pi(r)$. These are second-order
differential equations, and can be solved by the Runge-Kutta method,
or the Numerov method. In this work, the Runge-Kutta method is
adopted. Given the boundary values and a trial eigenvalue $E_{tr}$,
we can integrate those equations from $r=0$ outwards and from
$r=\infty$ (far away enough) inwards to a match point $r=r_m$, and
then connect the wave function at this point.  By integrating the
integrand that is $K_{tr}^{-*}(r)$ times Eq. (\ref{eqK}) in the range
$0<r<r_m-\epsilon$ and $r_m+\epsilon<r<\infty$ ($\epsilon$ is a
positive infinitesimal), we can get
\begin{equation}
E^2_{tr}=\int
K_{tr}^{-*}(r)(-\frac{d^2}{dr^2}+\frac{l(l+1)}{r^2}+M_K^2+\Pi(r))K_{tr}^-(r)dr/
\int K_{tr}^{-*}(r)K_{tr}^-(r)dr.
\end{equation}
However, we must keep in mind that $dK^-(r)/dr$ is discontinuous at
$r=r_m$ as long as $E_{tr}$ is still not the eigenvalue. This
discontinuity is used to adjust the eigenvalue in an iterative way
$E_{tr}^2=E_{tr}^2+\Delta E_{tr}^2$ with $\Delta E^2_{tr}$ being given as
\begin{eqnarray}
\Delta E^2_{tr}
 &\approx & \int_{r_m-\epsilon}^{r_m+\epsilon} -K_{tr}^{-*}(r)\frac{d^2K_{tr}^{-}(r)}{dr^2}dr/\int_{0}^{\infty}
 K_{tr}^{-*}(r)K_{tr}^{-}(r)dr \\
&\approx &
-K_{tr}^{-*}(r_m)[\frac{dK_{tr}^-(r_m+\epsilon)}{dr}-\frac{dK_{tr}^-(r_m-\epsilon)}{dr}]
/\int_{0}^{\infty} K_{tr}^{-*}(r)K_{tr}^{-}(r)dr,
\end{eqnarray}
where we have disregarded all infinitesimal terms in $\epsilon$ by
considering that the remaining term is from  the derivative of the
discontinuity which embodies the property of the Dirac $\delta$
function.

\end{document}